\documentclass[useAMS,usenatbib]{mn2e}
\usepackage{graphicx}
\usepackage{epstopdf}
 \def\thjet{\theta_{\rm j}}

\def \apj {ApJ\ }

\def \nat {Nature\ }

\title[]{The {\it Swift} short gamma-ray burst rate density: implications for binary neutron star merger rates}

\author[]{D.M. Coward$^{1,2}$\thanks{E-mail:david.coward@uwa.edu.au}, E.J. Howell$^{1}$, T. Piran$^{3}$, G. Stratta$^{4}$, M. Branchesi$^{5,6}$, \newauthor
 O. Bromberg$^{3}$, B. Gendre$^{4,8}$, R.R. Burman$^{1}$,  D. Guetta$^{7,8}$ \\
$^{1}$School of Physics, University of Western Australia, Crawley WA 6009, Australia\\
$^{2}$Australian Research Council Future Fellow\\
$^{3}$Racah Institute of Physics, The Hebrew University, Jerusalem 91904, Israel\\
$^{4}$ASI Science Data Center, via Galileo Galilei, 00044 Frascati (RM), Italy\\
$^{5}$DiSBeF - Universit\`a degli Studi di Urbino `Carlo Bo', I-61029 Urbino, Italy\\
$^{6}$INFN, Sezione di Firenze, I-50019 Sesto Fiorentino, Italy\\
$^{7}$Department of Physics and Optical Engineering, ORT Braude, P.O. Box 78, Karmiel, Israel\\
$^{8}$ INAF - Osservatorio Astronomico di Roma, Via Frascati 33, I-00040 Monteporzio Catone (Roma), Italy}

\begin{document}
\vspace{-5mm}
%\date{Accepted 1988 December 15. Received 1988 December 14; in original form 1988 October 11}

\pagerange{\pageref{firstpage}--\pageref{lastpage}} \pubyear{3002}

\maketitle

\label{firstpage}

\begin{abstract}
Short gamma-ray bursts (SGRBs) observed by {\it Swift} are potentially revealing the first insight into cataclysmic compact object mergers. To ultimately acquire a fundamental understanding of these events requires pan-spectral observations and knowledge of their spatial distribution to differentiate between proposed progenitor populations. Up to April 2012 there are only some $30\%$ of SGRBs with reasonably firm redshifts, and this sample is highly biased by the limited sensitivity of {\it Swift} to detect SGRBs. We account for the dominant biases to calculate a realistic SGRB rate density out to $z\approx0.5$ using the {\it Swift} sample of peak fluxes, redshifts, and those SGRBs with a beaming angle constraint from X-ray/optical observations. We find an SGRB lower rate density of $8^{+5}_{-3}$ $\mathrm{Gpc}^{-3}\mathrm{yr}^{-1}$ (assuming isotropic emission), and a beaming corrected upper limit of $1100^{+700}_{-470} $ $\mathrm{Gpc}^{-3}\mathrm{yr}^{-1}$. Assuming a significant fraction of binary neutron star mergers produce SGRBs, we calculate lower and upper detection rate limits of $(1-180)$ yr$^{-1}$ by an aLIGO and Virgo coincidence search. Our detection rate is similar to the lower and realistic rates inferred from extrapolations using Galactic pulsar observations and population synthesis.
\end{abstract}

\begin{keywords}
stars -- gamma-ray burst: individual -- gravitational waves -- techniques: miscellaneous -- stars: neutron
\end{keywords}

\section{Introduction}
It is generally accepted that `long' ($T_{90}>2$ s) \footnote{$T_{90}$ is the duration in which the cumulative counts are from 5\% to 95\% above background.} gamma-ray bursts (GRBs) are linked to the core collapse of massive stars (collapsars) \citep{woos93,paczy98,mac99}. For several cases, the GRBs are firmly associated with Type Ib/c
supernovae \citep[e.g.][]{hjorth03,stan03}. This
strongly suggests that they are linked to
the end point of massive stellar evolution. In contrast `short' ($T_{90}<2$ s) gamma-ray bursts (SGRBs) have a less certain origin.

The first breakthrough to understanding the nature of SGRBs was made in 2005 after the launch of the NASA {\it Swift} satellite \citep{geh04}. Prompt localizations and deep afterglow searches
yielded the first redshifts and investigations of their progenitor environments.
By late 2011, about three dozen SGRBs had been localized by {\it Swift}. Among them, about 50\% have optical detections and about one third have redshift determinations based on host galaxy spectroscopy.
%The reason for the lack of localizations is that SGRBs are comparatively fainter compared to long bursts.

Binary neutron star mergers (NS-NS) or neutron star--black hole (NS-BH) mergers are favoured as the progenitors for SGRBs, based on the association of some SGRBs with an older stellar population, as compared to long GRBs. Evidence for the origin of SGRBs in the final merger stage comes from the
host galaxy types \citep[e.g.][]{lrg05,z07}. Kicks imparted to NSs at birth will produce velocities of several hundred km s$^{-1}$, implying that binary inspiraling systems may
occur far from their site of origin. Fong, Berger \& Fox (2010) using Hubble Space Telescope observations to measure SGRB-galaxy offsets, find the offset distribution compares favourably with the predicted distribution for NS-NS binaries.

Despite this progress, unambiguous identification of burst types continues to be a problem. 
The various ambiguities have motivated several authors to re-define different classes of GRBs via a number of properties including: spectral features, associated supernova, stellar population, host galaxy, location in the host galaxy and progenitor type \citep{Zhang_GRBclass_06}. Some authors also suggest a third intermediate population of bursts based on the existing $T_{90}$ scheme \citep{Horvath_08} with a lower than average peak-flux distribution \citep{Veres_2010}.

Furthermore, work by \cite{brom12} shows that categorizing bursts as SGRBs based on $T_{90}$ is satellite dependent. They find that $2$ s $>T_{90}$ is statistically reasonable for BATSE bursts, but not accurate for {\it Swift} bursts. To quantify this uncertainty, for a 0.7-s burst they show that there is an equal probability of its being either long or short. 
%of a burst being an SGRB (non-collapsar) based on its {\it Swift} $T_{90}$.
 We account for this uncertainty, and its effect on the intrinsic SGRB rate, we scale our SGRB rate estimates using a refinement of these probabilities that incorporates a power law fit to the bursts within the {\it Swift} energy band.

Around 20\% of the SGRBs detected by \emph{Swift} have been followed by an extended emission lasting up to 100\hspace{1mm}s (hereafter SGRB-EE) \citep{norris_06,Perley_extendedSGRB_08} leading to suggestions that different progenitor types produce these bursts  \citep{norris_2011}. \cite{2008MNRAS.385L..10T} argue that SGRB-EE could be NS-BH mergers based on their galaxy off-sets. Other candidate systems include the birth of a rapidly rotating proto-magnetar produced via NS-NS merger or accretion-induced collapse of a white dwarf \citep{Metzger_2008,bucc11}. In the following, we have considered separately the normal SGRBs and the SGRB-EE to take this uncertainty in the progenitor nature into account.

SGRB observations could potentially play a more important role in constraining compact object merger rates on two fronts as more data become available. Firstly, as a direct consistency check of the binary merger rates inferred from binary pulsars and population synthesis. Secondly, as a constraint on the progenitor scenario and the engine that drives the emissions. Both are critical issues for solving the puzzle of the origin of SGRBs and their links to compact object mergers.

There are two popular but different methods employed for estimating NS-NS merger rates. The first uses extrapolation from
the observed sample of NS binaries infered from pulsar observations \citep{nps01,kal04}; the second uses population synthesis
simulations \citep{belcz07}, where several unknown model parameters are constrained by observations and others are
assumed from theory. \cite{lscrate} (hereafter Ab10) assessed the rates from these techniques, defining lower and realistic rate densities of 10 and 1000 $\mathrm{Gpc}^{-3}\mathrm{yr}^{-1}$ respectively. These rates are uncertain by $1-2$ orders of magnitude because of the small number of observed Galactic binary pulsars, and from poor constraints in population-synthesis models.

Within 5 years, co-ordinated gamma-ray, X-ray, optical and gravitational-wave observations may allow the strong gravity regime of the central engine of compact object mergers to be probed. Such `multi-messenger' observations provide the opportunity to probe these events across a vast energy spectrum, and to constrain the progenitor populations of SGRBs. Furthermore, co-ordinated optical and gravitational-wave searches may play an important role in confirming the first direct gravitational-wave observations of compact object mergers \citep{cow11}. It is becoming increasingly important to constrain the rate of compact object mergers and their proposed optical counterparts in the context of up-coming gravitational-wave searches.

The scarcity of SGRB observations and the poorly understood biases in the current data have made it challenging to calculate the SGRB rate density with meaningful uncertainties. Previous work using GRB flux-limited samples either assumes, or derives models for, the SGRB luminosity function and rate evolution of the sources; e.g. \cite{guetta06} find a SGRB rate density of $8-30$ $\mathrm{Gpc}^{-3}\mathrm{yr}^{-1}$, assuming isotropic emmision. Another study \citep{dietz11}, calculates a higher NS-NS merger rate of about $7800$ $\mathrm{Gpc}^{-3}\mathrm{yr}^{-1}$, but assuming the emissions are beamed with small opening angles.

In this study, we calculate a beaming-corrected SGRB rate density using the {\it Swift} sample of SGRB peak fluxes, redshifts and inferred beaming angles from X-ray observations.
For our GRB selection criteria we use the Jochen Greiner catalogue of localized GRBs (see Table 1) and select bursts indicated as short that have reliable redshifts up to 2012 April. From this selection of 9 bursts, we omit the burst GRB 090426, which is possibly linked to the death of a massive star based on its galaxy off-set and optical and x-ray emissions. We note also that although GRB 100816A has shown certain characteristics not consistent with a SGRB classification it cannot be clearly ruled out; additionally, its redshift of $z= 0.8$ will not influence our results significantly.

We avoid using a SGRB luminosity function, models for progenitor rate
evolution, and a beaming angle distribution, all of which have large uncertainties. Instead, we focus on observed and
measured parameters that take into account selection effects
that modify {\it Swift}'{\it s} detection sensitivity to SGRBs. Finally, we use our SGRB rate density estimates to infer a detection rate of binary NS mergers by Advanced LIGO (aLIGO) and Virgo interferomters. Despite the poor
statistics, this approach gives meaningful results and
can be followed up when a larger sample of SGRB observations becomes available.

\section{SGRB beaming constraints}\label{beam}
The currently favoured emission model for SGRBs is a compact object merger triggering an explosion causing a burst of collimated $\gamma$-rays \citep{elp+89,npp92, lrg05} powered by accretion onto the newly formed compact object.  The ultra-relavistic outflow is eventually decelerated by interaction with the interstellar medium to produce a fading X-ray and optical afterglow. After the jet Lorentz factor $\Gamma$ decreases to $\Gamma \sim \thjet^{-1}$, where $\thjet$ is the jet opening half angle, the afterglow becomes observable from viewing angles greater than $\thjet$.

Only two SGRBs have identifiable jet breaks \citep[see][and references therein]{cow11}: GRB 050709 ($\thjet\sim14\degr$ -- detected by the {\it HETE} satellite) and GRB 051221 $(\thjet\sim7\degr)$. These beaming angles are not necessarily agreed upon, mainly because the afterglow jet break times obtained from X-ray or optical light curves are notoriously difficult to identify at late times, and the emissions are generally fainter in X-rays and optical compared to those of long bursts. Furthermore, the inferred beaming angles are model dependent \citep{sari99} and sensitive to the circumburst environment. 

For the following rate density calculations, we employ the standard fireball model (e.g. Sari et al. 1999) to estimate the jet opening half angle using the minimum jet break times from X-ray observations:
\begin{equation}
\thjet=0.161[t_{j}/(1+z)]^{3/8}(n\eta/E_{\mathrm{iso}})^{1/8}\; \mathrm{rad},
\end{equation}
where $t_{j}$, the jet break time, is the time in days when the light curve decay index changes from $-1$ to $-2$. The parameter $n$ is the circum-burst density, in cm$^{-3}$, $\eta$ is the fraction of GRB energy observed in the prompt emission, and $E_{\mathrm{iso}}$ is the prompt isotropic equivalent energy in $10^{52}$ erg. We assume an average of $n=1$ cm$^{-3}$ and $\eta=0.1$ for our SGRB sample.

The circum-burst density for both long and short duration GRBs is quite scattered, and typically ranges from (0.001-10)cm$^{-3}$. \cite{kopac12} showed that $n$ for SGRBs is similar to the one derived for long bursts \citep{gendre06}, implying that the local density should be similar for both collapsars and mergers. Contrary to this, Fong et. al. (2012) argue that SGRBs are located in typically smaller circum-burst densities, supported by their constraint of $n=0.01-0.1$ cm$^{-3}$ for GRB 111020A. For definiteness, we use 0.01 cm$^{-3}$, to calculate lower limits on the beaming angle for SGRBs without a measured jet-break. Alternatively, for the SGRB-EE sample we use $n=1$ cm$^{-3}$, because the only confirmed SGRB-EE with a jet break, GRB 060614, is constrained to $n=10$ cm$^{-3}$ \cite{della06}. 

The difficulty of measuring jet breaks leads to a bias in obtaining jet angles favouring those bursts that have very bright X-ray/optical afterglows and/or relatively short jet break times. Hence, the true average of the SGRB jet angle distribution will be greater than that based on the very small sample used in this work. Nonetheless, given the lack of knowledge of the SGRB angle distribution, both theoretically and observationally, we choose to employ the small number of inferred SGRB beaming angles and lower limits from X-ray/optical observations.

To extend the very small measured beaming angle sample, we constrain the minimum beaming angles for those SGRBs that have X-ray afterglow light curves available up to at least 1 day from the trigger. The following five SGRBs: GRB 100816A, GRB 071227A, GRB 070714B, GRB 061006 and GRB 050724 satisfy this criterion.  We set a lower limit on the jet opening angle from the time when the {\it Swift} XRT monitoring stopped. We compute $E_{\mathrm{iso}}$,  from the literature, if available, or from the quoted fluences in the GCN circulars by estimating the corresponding $E_{\mathrm{iso}}$ in the 1--$10^4$ keV rest-frame energy range. For the GRB energy spectrum, we use the broken power law fit, termed the Band model \citep{band06}, or cut-off power law with spectral parameters where known. If the spectral parameters are unkown we simply rescale the fluence by $4\pi d_L^2/(1+z)$, where $d_L$ is the luminosity distance.

\begin{table}
 \begin{tabular}{@{}lccccc}
\hline
\hline
SGRB  & $T_{90}$ &  $\thjet$   &  $z$   &  peak flux & ref\\
      & (s)  & (deg) &  & (ph s$^{-1}$ cm$^{-2}$) &\\
\hline
\hline
101219A$*$  & 0.6 &  - &  0.718  & 4.1& (1)   \\
%100816A$*$  & 2.9 &   $>12$$^\dagger$ &  0.803  & 10.9 & (2)   \\
%100724A$*$  & 1.4 &   $>14$$^\dagger$ &  1.288  & 1.9 & (3)   \\
100117A$*$  & 0.3 &   - &  0.92  & 2.9 & (2)   \\
090510A$*$   & 0.3 &   - &  0.903  & 9.7 & (3)   \\
080905A\boldmath {$^a$}  & 1.0 &   - &  0.122  & 6.0 & (4)   \\
070724A  & 0.4 &   $>11$$^\dagger$ &  0.457  & 2.0 & (5)   \\
061217A   & 0.2 &   - &  0.827  & 2.4 & (6)  \\
%060502B  & 0.1 &  - &  0.287  & 8.5 & (8)   \\
051221A$*$ & 1.4  &   7\boldmath {$^b$} &  0.547  & 12.0  & (7)  \\
050509B  & 0.73 &   - &  0.225  & 3.7 & (8)   \\
\hline
SGRB-EE & & & & &\\
\hline
071227A & 1.8   &   $>16$$^\dagger$ &  0.383  & 4.4 & (9)   \\
070714B & 64  &   $>6$$^\dagger$ &  0.923  & 10.0 & (10)  \\
061210 & 85   &  $> 12$$^\dagger$ &  0.409  & 62.9 & (11)   \\
061006 &129  &   $>$6$^\dagger$ &  0.437  & 15.8 & (12)  \\
060614 & 109   & 12\boldmath {$^c$} & 0.125 & 25.0 & (13) \\
050724 & 96   &   $>$25$^\dagger$ &  0.258  & 12.8 & (14),(15)   \\
\hline
\hline

\end{tabular}
\caption[]{SGRB peak fluxes, $T_{90}$ and redshifts taken from the \emph{Swift} online catalogue and http://www.mpe.mpg.de/$\sim$jcg/grbgen.html used to calculate Poisson rates. We use the 20-ms peak photon fluxes from the BAT2 catalogue where possible -- those marked by $*$ are 1-s peak photon fluxes. $^\dagger$ We set a lower limit on the jet opening angle from the time when the {\it Swift} XRT monitoring stopped.\\
{\boldmath $^a$}The proposed host galaxy at $z = 0.1218$ for GRB 080905A (Rowlinson et al. 2010) is a strong outlier to the Yonetoku relation \citet{2004ApJ...609..935Y}. A redshift $z > 0.8$ would make it consistent \citep{gruber12}.\\
{\boldmath $^b$} The detection of a jet break was observed for GRB 051221A in X-rays (Soderberg et al. 2006) at about 5 days post-burst, or $\theta\sim7$\degr.\\
 {\boldmath $^c$} Jet break identified at 1.4 days post-burst, circum-burst density $n=10$ cm$^{-3}$, and $E_{{\mathrm iso}}=2.5\times10^{51}$ ergs.
\newline References--(1) GCN 11518,  (2) \citet{fong_2011}, (3) \citet{Ackermann_2010}, (4) \citet{Rowlinson_2010}, (5) \citet{0004-637X-704-1-877}, (6) \citet{berger_07}, (7) \citet{Sod06},
(8) \citet{2005Natur.437..851G}. The SGRBs below the line are classified as SGRB-EE -- (9) \citet{2010A&A...521A..80C}, (10) \citet{0004-637X-698-2-1620}, (11) \citet{berger_07}, (12) \citet{berger_07}, (13) \cite{della06}, (14) \citet{berger_2005}, (15) \citet{grup06}. }
\label{table_swift_z_data6}
%\end{centering}
\end{table}

\section{Empirical SGRB rate model}
A Poisson GRB rate can be estimated from small number statistics using $V_{\mathrm{max}}$ \citep{piran92,cohen95}, where $V_{\mathrm{max}}$ is the maximum volume within which a SGRB with observed peak flux, $F_p$, could be detected for a given satellite detector with detection sensitivity $F_{\mathrm{Lim}}$. While $V_{\mathrm{max}}$ has been used previously  \cite[e.g.][]{guetta07} to estimate rates of low-luminosity GRBs \cite[see also][]{cow05}, we extend the rate estimate method to account for biases in the {\it Swift} GRB redshift and peak flux sample.

Firstly, the low energy detection bandwidth of {\it Swift} (15--150 keV) in comparison with BATSE's 50--300 keV results in a bias against SGRBs which typically have harder emissions. Secondly, the {\it Swift} detection threshold is not simply defined by the detector sensitivity, but by a complex triggering algorithm. Both these effects manifest as a bias against {\it Swift} detecting SGRBs; i.e. a smaller proportion of bursts has been detected by
{\it Swift} ($\sim$10\%). The latter effect results from the detection process employed by BAT, the \emph{Swift} coded-aperture mask $\gamma$-ray detector. In addition to requiring an increased photon count rate above background (the sole triggering criterion used for BATSE), BAT employs a second stage in which an image is formed by accumulating counts for up to 26\,s \citet{band06}.

We attempt to crudely correct for these biases by using the observed SGRB rate from BATSE as a rate calibration for the \emph{Swift} SGRB rate. Because BATSE operated at different energy thresholds and trigger sensitivities, for consistency we take
all BATSE SGRBs with 64-ms peak flux when the trigger threshold was set to 5.5 $\sigma$ in the 50--300 keV energy range (total live operation time of 3.5 years). This yields 32 SGRBs sr$^{-1}$ yr$^{-1}$, assuming an effective BATSE FoV of $\pi$ sr \citet{band03}. The ratio of the rate of BATSE to \emph{Swift} SGRBs, $R_{B/S}=6.7$, is used to calibrate the observed {\it Swift} SGRB rate to estimate an intrinsic SGRB rate.

Because SGRBs occur over a short duration, it is more difficult (compared to long bursts) to produce a significant signal above background.  Hence instead of using the theoretical BAT sensitivity of $F_{\mathrm{Lim}}=0.4$ ph\,s$^{-1}$ cm$^{-2}$, we employ a flux limit of 1.5 ph\,s$^{-1}$ cm$^{-2}$, using the smallest 20-ms peak flux from the SGRB data.

%For bursts in which a 20\,ms peak flux is not available we use the 1\,s peak fluxes of the \emph{Swift} online catalogue\footnote{http://swift.gsfc.nasa.gov/docs/swift/archive/grb\_table} for which the equivalent value is 1.67 ph\,s$^{-1}$.

Another detector-dependent selection effect arises from the limited energy bandpass of {\it Swift}, i.e. (15--150) keV. What is measured by {\it Swift} is not the bolometric peak flux, but a detector response and source-spectrum dependent peak flux. We define the bolometric isotropic-equivalent luminosity correction factor \citep{imerit09}:
\begin{equation}
C_{\mathrm{det}}(e_1,e_2) \equiv  \frac{\int_{E_1}^{E_2} E N(E) {\rm d}E}{\int_{e_1}^{e_2} E N(E) {\rm d}E}\;,
\end{equation}
where $[E_1=1, E_2=10000]$ keV spans the bolometric gamma-ray spectrum, and $[e_1, e_2]$ is the sensitivity band of {\it Swift}, i.e. 15 to 150 keV. For the source energy spectrum, $N(E)$, we employ the Band function \citep{band06}, with spectral indices $\alpha=-1$, $\beta=-2.3$ and a rest-frame power-law break energy of 511 keV.

In addition, for high redshift sources, a higher energy component of the source spectrum is redshifted into the sensitivity band of the detector. The following factor, $k(z)$, accounts for the downshift of $\gamma$-ray energy from the burst to the observer's reference frame:
\begin{equation}
k(z) \equiv  \frac{\int_{e_1}^{e_2} E N(E) {\rm d}E}{\int_{(1+z)e_1}^{(1+z)e_2} E N(E) {\rm d}E}\;.
\end{equation}

Taking into account the sensitivity reduction and k-correction, the SGRB all-sky rate can be inferred from the flux-limited SGRB sample in Table \ref{table_swift_z_data6}. We calculate the maximum distance $d_{\mathrm{max}}$, with corresponding redshift $z_{\mathrm{max}}$, that a burst at luminosity distance $d_L$ could be detected given \emph{Swift's} sensitivity $F_{\mathrm{Lim}}$:
\begin {equation}
d_{\mathrm{max}} =\sqrt{ \frac{F_p k(z)}{ F_{\mathrm{Lim}}k(z_{\mathrm{Lim}})}}d_L(z)\;,
\end{equation}
where $F_p$ is the observed peak flux and $k(z_{\mathrm{Lim}})$ is the k-correction for a burst at the maximum limiting distance of detection for the sample. Given the large uncertainty in the SGRB luminosity function, we use the largest redshift from our sample, $z_{\mathrm{Lim}}=0.92$. We note that the bolometric correction, $C_{\mathrm{det}}(e_1,e_2)$, cancels out in the above equation, because it is independent of redshift.

The corresponding maximum SGRB detection volume for each burst is defined as
\begin {equation}
V_{\mathrm{max}} = \int_{0}^{z_{\mathrm{max}}} \frac{dV}{dz} dz\;,
\end{equation}
where the volume element factor, $dV/dz$, and luminosity distance, $d_L(z)$, are calculated using a flat-$\Lambda$ cosmology with $H_{\mathrm 0}$ = 71 km s$^{-1}$ Mpc$^{-1}$, $\Omega_M$ = 0.3 and $\Omega_\Lambda$ = 0.7.  This estimate (Eq. 5) is valid as long as the rate of SGRBs does not change significantly over the range $(0,z_{\mathrm{max}})$. Since the detection range of SGRBs is quite small and typically $z < 1$ this approximation is valid.

To calculate the intrinsic rate of SGRBs requires accounting for the beaming angle, $\thjet$, of the jetted burst. Equation (\ref{eq3}) expresses the beaming factor used to correct for the unobserved SGRBs that are not detected because the jet is misaligned with the detector:
\begin{equation}
B(\thjet) =  [1- \mathrm{cos} (\thjet)]^{-1}\,.
\label{eq3}
\end{equation}

To account for the fact that only a fraction of observed SGRB have measured redshifts, we scale the rate density by the ratio of {\it Swift} bursts with redshift to those without redshifts, $F_r$. We use $F_r \approx 8/39$ and $6/12$  for SGRB and SGRB-EE respectively. The time span encompassing all observations, $T\sim6$ yrs, is defined by the start of {\it Swift} observations to the time of the most recent SGRB in the sample and we account for the fractional sky coverage of {\it Swift}, $\Omega\approx0.17$. To account for {\it Swift}'{\it s} reduced sensitivity for detecting SGRBs relative to BATSE, we use the ratio of the BATSE to {\it Swift} SGRB detection rate, which we approximate as $R_{B/S}=6.7$. This converts the {\it Swift} SGRB rate to an intrinsic SGRB rate. We point out that this correction applies only to SGRB, and not SGRB-EE, because {\it Swift} is more sensitive to these longer emission bursts.

Finally, we calculate the probability of a GRB to be a non-collapsar, P$_{{i}(T_{90}; {\rm P}_{\rm L})}$ \citep{brom12}, based on its $T_{90}$ and a power-law fit to the gamma-ray spectrum in the $15-150$ keV band. The study was refined \citep{brom12b,brom12c} to show that a GRB with a hard power-law photon index (P$_{{\rm L}}<1.15$) has a larger chance of being a non-collapsar than a GRB with a soft photon index at the same $T_{90}$. We note that these probabilities do not influence the magnitude of the calculated rate densities significantly in this work.

Combining all detection parameters yields the Poisson SGRB rate density of a single ($i$th) burst, and the total rate density for $n$ bursts:
\begin {equation}\label{Rate}
R_{\mathrm{SGRB}} = \sum_{i}^{n}  \frac{1}{V_{\mathrm{i(max)}}} \frac{1}{F_r} \frac{1}{T} \frac{1}{\Omega} R_{B/S}  B_i(\thjet) {\rm P}_{{i}(T_{90}; {\rm P}_{\rm L})}.
\end{equation}

SGRBs may track the star formation rate history (SFR), albeit with a time delay. The $z_{{\rm max}}$ that we derive for the most significant contributions to the rate density range from $z= 0.2-0.4$. To investigate the effect of including SFR evolution, we recalculated a local rate density using an integrated differential rate equation using several different SFR models and volumes bounded by $z_{{\rm max}}$. We find differences of a factor of about 2 between the local rate densities calculated using equation \ref{Rate} and an integrated differential rate model. Also, there are discrepancies at small-$z$ using different SFR models, highlighting that the models are not good fits at small $z$. Given that the uncertainties in the small $z$ SFR are of the same scale as the change in the SFR from $z=0-0.3$, e.g. see figure 10 in Reddy \& Steidel (2009), we do not include an evolution of the SGRB rate density in space.  

We calculate lower rate estimates assuming no beaming. Upper rates are estimated using the observed beaming angle constraints for each burst, and if a beaming angle estimate is unavailable, we use the smallest inferred jet angle from the data, $\thjet\sim7\degr$. Given that the inferred average beaming angle of the jet is biased towards SGRBs with short jet-break times (see \S \ref{beam}), our upper rate limit estimate
is at the limit of plausibility.

We use the above SGRB rate density to infer a detection rate of binary NS mergers by advanced gravitational-wave interferometers. For aLIGO and Virgo interferometer sensitivities, the horizon distance, $D_h$ (all sky locations and orientation averaged over) for optimal detection of a NS-NS merger in a coincidence search is about 340 Mpc (cosmological redshift not included).  For a direct comparison with Ab10, the detection rate is computed considering the noise power spectral spectral density of a single interferometer, with $D_h=197$ Mpc.  
Given the uncertainty in the beaming angle distribution, we define an optimal detection rate as a function of $\thjet$ using the SGRB lower rate estimate, i.e. from $\thjet=90\degr$, scaled by $B(\thjet)$ and the Euclidean volume: 
\begin{equation}
R(\thjet) = \frac{4\pi}{3} D_h^3 R_{\mathrm{Low}}B(\thjet).
\label{gw}
\end{equation}

\begin{table}
 \begin{tabular}{@{}lcccc}
\hline
\hline
      SGRB  & P$_{{i}(T_{90}; {\rm P}_{\rm L})}$  &     lower rate   &   upper rate  \\
         &   &  Gpc$^{-3}$ yr$^{-1}$             &  Gpc$^{-3}$ yr$^{-1}$ \\
\hline
\hline
%101219A  &  0.79 & 0.031 &  5.6\\
%100816A  &  0.51 & 0.0099 &  0.45\\
%100724A  &  0.72 & 0.021 &  0.7\\
%090510   &  0.89 & 0.016 &  2.9\\
%080905A  &  0.72 & 3.6 &  670\\
%070724A  &  0.84 & 0.58 &  110\\
%061217   &  0.89 & 0.15 &  28\\
%060502B  &  0.72 & 0.32 &  58\\
%051221A  &  0.72 & 0.021 &  2.8\\
%050509B  &  0.79 & 1.5 &  270\\

101219A  &  0.79 & 0.039 &  5.3$^\star$\\
%100816A  &  0.24$^\dagger$  & 0.005 &  0.2\\
100117A  &  0.89 & 0.039 &  5.2$^\star$\\
090510   &  0.89 & 0.02 &  2.7$^\star$\\
080905A\boldmath {$^a$}  &  0.72 & 4.9 &  660$^\star$\\
070724A  &  0.84 & 0.77 &  140\\  
061217   &  0.89 & 0.2 &  27$^\star$\\
051221A  &  0.72 & 0.026 &  3.5\\
050509B  &  0.79 & 2 &  270$^\star$\\   

\hline
Total rate & &   $8^{+5}_{-3} $ & $1100^{+700}_{-470} $\\
\hline
\hline
SGRB-EE & &\\
\hline
071227   &  &  0.038 &  0.97\\
070714B  & &   0.0036 &  0.66\\
061210   & &  0.0034 &  0.16\\
061006   &   & 0.0088 &  1.6\\
060614   &  & 0.073 &  3.3\\
050724   &  & 0.03 &  0.32\\
\hline
Total rate &  & $0.16^{+0.15}_{-0.088} $ & $7.1^{+6.9}_{-4} $\\
\hline
\hline
\end{tabular}
\caption[]{The beaming-corrected SGRB rate densities with Poisson uncertainties using the observed constraints on $\thjet$, and scaled by the probability P$_{{i}(T_{90}; {\rm P}_{\rm L})}$, except GRB 100816A, which uses just the $T_{90}$.
Lower rate estimates assume isotropic emission, and upper rates use the observed beaming angle constraints shown in Table 1, or the smallest observed beaming angle in the sample$^\star$, $\thjet\approx7\degr$. We calculate the SGRB-EE rate density separately, noting that $T_{90}$ is generally much larger than 2 s, as shown in Table 1.\\
 \boldmath {$^a$} Because of the importance of GRB 080905 for the rate density, and its uncertainty (see Table 1. for caveats), we also calculate total rates excluding this burst i.e. $(3^{+2}_{-1}-500^{+340}_{-220})$ Gpc$^{-3}$ yr$^{-1}$.} 
\label{table2}
\end{table}

\section{Results}
We have focussed on the sensitivity of {\it Swift} for detecting SGRBs, and highlight several important issues. The complex triggering algorithm of the BAT is biasing detection against SGRBs, and this should be considered when attempting to estimate an intrinsic rate density of SGRBs. The rate estimates using $V_{\mathrm{max}}$ are sensitive to the flux limit of the detector, but are mostly invariant to the bolometric and k-correction because these effects cancel out. We also note that SFR evolution may increase the calculated SGRB rate density by $\sim2$ in the small-$z$ regime.

With the above caveats, Table 2 shows the rate densities of SGRB and SGR-EE using equation (\ref{Rate}) applied to each burst, with the observed constraints on $\thjet$. The total lower and upper rate densities for SGRBs are $8^{+5}_{-3} $ (assuming isotropic emission) and $1100^{+700}_{-470} $  $\mathrm{Gpc}^{-3}\mathrm{yr}^{-1}$ (assuming beaming--see Table 2) respectively, where the errors are the 95\% Poisson statistics \citep{1986ApJ...303..336G}.  The SGRB rate density is dominated by GRB 080905A, which was relatively faint and nearby, with a $z_{\mathrm{max}}\approx0.3$. Because of the uncertainty in the redshift of this burst, we also calculate a rate density excluding GRB 080905A, and find total and upper rate densities of $(3^{+2}_{-1}-500^{+340}_{-220})$ Gpc$^{-3}$ yr$^{-1}$. To test the robustness of the total rate density given that only a few SGRBs contribute significantly to the sum, we used ``jackknife" on the summed SGRB rates in Table 2. The standard error was found to be $\pm 240$ Gpc$^{-3}\mathrm{yr}^{-1}$, which is less than the Poisson counting error and the beaming angle uncertainty.

For SGRB-EEs, we find a corresponding rate density  of $0.16^{+0.15}_{-0.088} $ and $7.1^{+6.9}_{-4} $ $\mathrm{Gpc}^{-3}\mathrm{yr}^{-1}$ respectively. We note that these rates assume $n=1$ cm$^{-3}$. If a smaller circum-burst density similar to the SGRB is used, the rates will increase by a factor of about 2. It is interesting that the SGRB-EE rate density from this work is similar to the expected BH-NS merger rates from population synthesis models (Ab10).

Assuming a signicant fraction of binary NS mergers produce SGRBs, it is interesting to compare these rate estimates to the NS-NS merger rate density estimated from other work. Our lower
 rate density, which assumes isotropic emission, is comparable to the isotropic emission estimates of \cite{guetta06}, i.e. $(8-30)$  $\mathrm{Gpc}^{-3}\mathrm{yr}^{-1}$. Our upper limit is comparable to estimates of beamed emission, $(240-1500)$ $\mathrm{Gpc}^{-3}\mathrm{yr}^{-1}$. Compared to Ab10, our limits of $(7-1200)$ $\mathrm{Gpc}^{-3}\mathrm{yr}^{-1}$ are similar to their low-realistic rate densities $(10-1000)$ $\mathrm{Gpc}^{-3}\mathrm{yr}^{-1}$.

Finally, we estimate a plausible detection rate by aLIGO/Virgo using the SGRB rate density limits and equation (\ref{gw}). The corresponding lower and upper detection rates using the SGRB rates from Table 2 are $(0.2-40)$ yr$^{-1}$ for a single interferometer with $D_h=197$ Mpc. Ab10 give a detection rate for aLIGO of (0.4, 40, 400) yr$^{-1}$ for low, realistic and high detection rates respectively. In our study, an optimal coincidence search at aLIGO/Virgo sensitivities, with $D_h=341$ Mpc, the detection rate increases to $(1-180)$ yr$^{-1}$.

Figure 1 plots the detection rate for a single aLIGO interferometer ($D_h=197$ Mpc), and an optimal coincidence search with aLIGO/Virgo ($D_h=341$ Mpc), as a function of SGRB beaming angle using equation (\ref{gw}). We use the SGRB rate density $8$ $\mathrm{Gpc}^{-3}$ $\mathrm{yr}^{-1}$ (see Table 2) that assumes isotropic emission. 

\begin{figure}
%\centering
\includegraphics[scale=0.53]{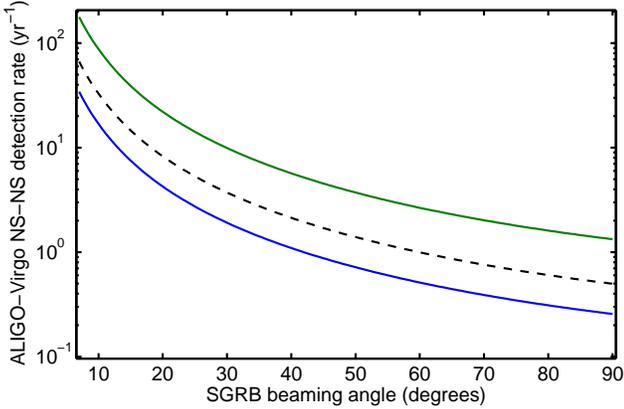}
\caption{The lower curve plots equation (\ref{gw}), the detection rate of binary NS mergers by a single aLIGO interferometer ($D_h=197$ Mpc) as a function of SGRB beaming angle, using $R_{\mathrm{Low}}=8$ $\mathrm{Gpc}^{-3}\mathrm{yr}^{-1}$ (see Table 2). The upper curve assumes a coincidence search with aLIGO and Virgo interferometers with $D_h=341$ Mpc. Both horizon distances used to calculates detection rates are angle averaged over all binary orientations. The dashed curve is the same as the upper curve but using the upper rate excluding GRB 080905 (see Table 1. for caveats).
} \label{fig1}
\end{figure}

Our detection rate is similar to the lower and realistic rates based on population synthesis simulations using Galactic pulsar observations. The rates are also compatible with those of Guetta \& Piran (2006), who employ a luminosity function and SGRB rate evolution.

%Although this work does not assume knowledge of SGRB progenitors, it is interesting to compare the calculated SGRB rate density with that from current estimates of compact object merger rates. We consider the most plausible progenitor scenarios and rate estmates within the framework of current modelling:
%\begin{enumerate}
%\item {\bf A significant fraction of binary NS mergers produce SGRBs}:\newline
%
%\item {\bf A very small fraction of binary NS mergers produce SGRBs}:\newline
%
%\item  {\bf A significant fraction of binary NS mergers produce SGRB-EEs}:\newline

%Some models suggest that the SGRB-EE could be powered by a nascent proto-neutron star magnetar driving relativistic winds at late times \citep{bucc11}.
%\item  {\bf A significant fraction of BH-NS mergers produce SGRB-EEs}:\newline

%\item  {\bf A significant fraction of accretion induced collapses of white dwarfs produce SGRB-EEs}:\newline An accreting white dwarf collapses to form a nascent proto-neutron star/magnetar driving relativistic winds at late times \cite[see e.g.][]{Metzger_2008,bucc11}. Hence, the calculated SGRB-EE rate could represent a magnetar formation rate for this particular evolutionary channel.

%\end{enumerate}

\section{Discussion}\label{discuss}
It has been suggested that the observed SGRB population is presently not a useful constraint on rate estimates for compact binary mergers (Ab10). The justification of this argument stems from the fraction (if any) of binary NS mergers that will produce an SGRB. Another issue is that SGRB observations by {\it Swift} and subsequent follow-up in optical are affected by selection effects and uncertainty in the beaming angles. If SGRB/SGRB-EE have an intrinsic average beaming angle of $\sim 50\degr$, then the jet-break times may not occur until weeks after the burst, when the afterglow is too faint in the optical.

The SGRB rate density depends on the spatial-distribution of the sources, but because redshift measurement depends on a bright afterglow, the spatial distribution is biased to the brighter bursts that are nearby \citep{cow08,cow09,imerit09}. Hence, for the majority of bursts with a faint afterglow, redshift measurement is more difficult. We attempt to crudely correct for this bias by boosting the calculated rate by the fraction of bursts without measured redshift to those with redshift, assuming the missing redshifts follow the same distribution as the observed redshifts. This assumption is reasonable (in this work) because the rate density is dominated by those faint bursts in the smallest $V_{\mathrm{max}}$.  

Hence, there are two main sources of selection effects and biases. Firstly, there are the satellite detectors, exemplified by the significant difference between BATSE and {\it Swift} SGRB detection efficiencies. Secondly, there is the problem of obtaining a redshift either from the host galaxy or the afterglow itself (as highlighted above). We have shown that these detection biases should be considered when attempting to use the current (and future) SGRB detections for constraining rate evolution, rate densities and linking SGRBs to binary NS mergers.

The binary NS gravitational-wave detection rate estimates are based on calculating an intrinsic SGRB rate density using {\it Swift} localized bursts, taking into account dominant selection effects. This approach, based on observational data is very different from that based on Galactic binary pulsar observations and modelled population synthesis.
In the latter, Ab10 use the observed Galactic binary pulsar population 
and extrapolate a NS merger rate density out to the aLIGO and
Virgo detection horizon. Conversely, our approach avoids this
extrapolation because it is essentially an observed rate extending
well beyond the average sensitivity distances of the upcoming
gravitational-wave searches for compact binaries (about 300 Mpc or $z =
0.07$ for aLIGO and Virgo interferometers) and, moreover,
a significant fraction of SGRBs are observed in association 
with evolved stellar populations. 

In conclusion, the upcoming gravitational-wave detection era will be fundamental for resolving the SGRB--binary NS merger connection, since an unequivocal association between SGRBs and binary NS mergers will only be possible via coincident gravitational-wave and electromagnetic observations. Ultimately, a comparison beteween the SGRB rate density and the gravitational-wave detection rate will help constrain the fraction of binary NS mergers that give rise to SGRBs and the SGRB beaming angle distribution.

%\vspace{-5mm}
\section*{Acknowledgments}
D.M. Coward is supported by an Australian Research Council Future Fellowship. G. Stratta and B. Gendre acknowledge support from ASI through ASI grants I/009/10/0. T. Piran and O. Bromberg acknowledge support by an Advanced ECR grant. We thank the referee for a careful review with several useful suggestions.

%\vspace{-5mm}
\label{lastpage}
\end{document}